\documentclass[letterpaper,twocolumn,10pt]{article}     %
\usepackage{usenix-2020-09}                             %
\usepackage[english]{babel}
\usepackage{balance}
\usepackage{blindtext}
\usepackage{xcolor,colortbl}
\usepackage{lipsum} %
\usepackage{enumitem}
\usepackage[linesnumbered,norelsize,ruled,vlined]{algorithm2e}
\usepackage{subcaption}
\usepackage{algpseudocode}
\usepackage{comment}
\usepackage{amsmath}                                   %
\usepackage{cleveref}
\usepackage{url}
\usepackage{etoolbox} %
\usepackage{microtype}
\usepackage{textcomp}
\usepackage{mathtools}
\usepackage{bm}
\usepackage{tikz}
\usepackage{float}
\usepackage{tabularx}
\usepackage{pifont}%
\usepackage{listings}
\usepackage{multirow}
\usepackage{makecell}
\usepackage{stfloats}
\usepackage{titlesec}
\usepackage{titling}                                   %
\usepackage{setspace}                                  %
\usepackage{tcolorbox}
\usepackage{upgreek}
\usepackage{placeins}

\usepackage{listings}
\usepackage{xcolor}

\definecolor{codegray}{rgb}{0.5,0.5,0.5}
\definecolor{codepurple}{rgb}{0.58,0,0.82}
\definecolor{backcolour}{rgb}{0.95,0.95,0.92}
\definecolor{functioncolor}{rgb}{0.1, 0.5, 0.1} %
\definecolor{lightorange}{RGB}{255,247,230}
\definecolor{verylightgray}{rgb}{0.95,0.95,0.95}
\definecolor{darkgreen}{HTML}{008000}

\lstdefinestyle{mystyle}{
    backgroundcolor=\color{verylightgray},   
    commentstyle=\color{codegray},
    keywordstyle=\color{blue},
    numberstyle=\tiny\color{codegray},
    stringstyle=\color{codepurple},
    basicstyle=\ttfamily\footnotesize,
    breaklines=true,
    captionpos=b,
    numbers=left,
    numbersep=5pt,
    showspaces=false,
    showstringspaces=false,
    showtabs=false,
    tabsize=2,
    escapeinside={(*@}{@*)}, %
}

\newcommand{\paraspace}{\vspace{0.05in}}
\newcommand{\para}[1]{\paraspace\noindent\textbf{#1}}

\newcommand{\hideme}[1]{}
\SetNlSty{bfseries}{\color{black}}{}

\SetCommentSty{mycommfont}

\usepackage{hyphenat}

\def\name{Mercury}

\def\llama{llama.cpp}

\def\bestimprove{53.4\%}

\titlespacing*{\section}{0pt}{7pt plus 3pt minus 3pt}{3pt plus 3pt minus 2pt}
\titlespacing*{\subsection}{0pt}{4pt plus 3pt minus 2pt}{1pt plus 3pt minus 1pt}

\titlespacing*{\subsubsection}{0pt}{4pt plus 3pt minus 2pt}{1pt plus 3pt minus 1pt}
\titleformat{\subsection}{\normalfont\bfseries}{\thesubsection}{1em}{}
\titleformat{\subsubsection}{\normalfont\bfseries}{\thesubsubsection}{1em}{}

\newenvironment{denseitemize}{
\begin{itemize}[noitemsep, topsep=1pt, leftmargin=1.5em]
}{\end{itemize}}

\date{}         %

\begin{document}
\sloppy

\title{\Large\bf{\name}: QoS-Aware Tiered Memory System \vspace{-2ex}}
\author{\normalsize Jiaheng Lu*, Yiwen Zhang*, Hasan Al Maruf$^\ddagger$, Minseo Park$^\ddagger$, Yunxuan Tang, Fan Lai$^\mathsection$, Mosharaf Chowdhury \\\normalsize
\textit{University of Michigan $^\ddagger$AMD $^\mathsection$UIUC}
}
\maketitle
\def\thefootnote{\textbf{*}}\footnotetext{Equal contributions.}\def\thefootnote{\arabic{footnote}}
\pagenumbering{arabic}  %
\begin{abstract}
Memory tiering has received wide adoption in recent years as an effective solution to address the increasing memory demands of memory-intensive workloads.
However, existing tiered memory systems often fail to meet service-level objectives (SLOs) when multiple applications share the system because they lack Quality-of-Service (QoS) support.
Consequently, applications suffer severe performance drops due to \emph{local memory contention} and \emph{memory bandwidth interference}.

In this paper, we present Mercury, a QoS-aware tiered memory system that ensures predictable performance for coexisting memory-intensive applications with different SLOs.
Mercury enables per-tier page reclamation for application-level resource management and uses a proactive admission control algorithm to satisfy SLOs via per-tier memory capacity allocation and intra- and inter-tier bandwidth interference mitigation.
It reacts to dynamic requirement changes via real-time adaptation.
Extensive evaluations show that Mercury improves application performance by up to 53.4\% and 20.3\% compared to TPP and Colloid, respectively.

\end{abstract}

\section{Introduction}
\label{sec:introduction}

With the increasing memory demands of datacenter applications, tiered memory systems have been widely adopted to replace DRAM-only systems~\cite{canvas, hemem, nimble, tmts, vtmm, pond}.
Recent tiered memory systems are enabled by Compute Express Link (CXL)~\cite{cxl}, a high-speed interconnect standard that allows memory capacity to grow by attaching memory to the CPU without losing nanosecond-scale memory access latency. %

The increased memory capacity enables deployments of more memory-intensive applications that fall into two broad categories: (1) \textit{latency-sensitive (LS)} applications, such as in-memory key-value store~\cite{memcached, redis}, that require low-latency memory access, and (2) \textit{bandwidth-intensive (BI)} applications, such as large-memory machine learning (ML) models (e.g., long-context language models or recommendation models~\cite{dlrm, llama}), that require sustained memory bandwidth. %
In production environments, running a single application on a tiered memory system wastes both memory capacity and bandwidth~\cite{tpp, tmts, pond, vtmm, tmc, atlas}.
As a result, multiple memory-intensive applications often share the tiered memory system to maximize resource utilization and improve total cost of ownership (TCO).

Existing research on tiered memory systems primarily focuses on page temperature monitoring and efficient page migration to better utilize memory resources (i.e., fast-tier DRAM)~\cite{hemem, nimble, autotiering, tpp, thermostat, HeteroOS, memstrata}.
However, these solutions optimize a single application running on a single server. 
They lack Quality-of-Service (QoS) support and cannot react to applications with different service level objectives (SLOs).
In particular, existing tiered memory systems suffer from two sources of performance unpredictability (\S\ref{sec:moti-exp}).
First, multiple applications can \textit{contend for local memory} (i.e., fast-tier memory), and the ones with hotter memory get more resources. %
As a result, a low-priority application may acquire more local memory than a high-priority application, leading to \textit{priority inversion}.
Second, high memory bandwidth generated by BI applications can degrade the performance of coexisting LS applications on the same tier or even across tiers -- we refer to these phenomena as \textit{intra- and inter-tier interference}, respectively.
To the best of our knowledge, no existing solutions have systematically tackled both of these QoS challenges in a tiered memory system, including recent proposals on QoS support for tiered memory~\cite{tmts, colloid}.

We observe that providing predictable performance for applications with different SLOs in the presence of local memory contention and memory bandwidth interference requires addressing the following fundamental challenges:
\begin{denseitemize}
    \item \emph{Efficient resource tracking and control:} The memory management subsystems in existing operating systems (OSes) are not well designed for tiered memory. 
        Hence, efficiently tracking and controlling memory resources on each memory tier is challenging.

    \item \emph{Optimal memory allocation:} Determining the amount of memory to allocate to each application in each tier while maximizing resource utilization is non-trivial. 
        For instance, migrating some local memory of a BI application with a relaxed SLO to CXL memory can save local memory for another LS application.
        However, excessive migration may cause inter-tier interference, leading to SLO violations.

    \item \emph{Dynamic workload adaptation:} Because an application's workload can change after deployment, we need an efficient approach to performance monitoring and real-time adaptation.
\end{denseitemize}

In this paper, we present \name{}, a QoS-aware tiered memory system that provides predictable performance for memory-intensive applications.
\name{} combines several key design ideas to overcome the aforementioned challenges.
It achieves QoS-awareness in tiered memory by enabling \textit{per-tier page reclamation} inside Linux kernel. This allows \name{} to perform efficient application-level resource management while still leveraging existing page migration designs. 
At the core of \name{}'s design is a novel admission control algorithm tailored for tiered memory.
It proactively allocates the right amount of resources for applications to satisfy SLOs while mitigating intra-tier and inter-tier bandwidth interference.
\name{} also performs real-time adaptation to unpredictable memory and bandwidth changes, preventing sudden load surges while maximizing the number of applications meeting their SLOs based on priority.

We evaluate \name{}'s effectiveness in providing QoS using real-world applications.
\name{} can closely track SLOs at different memory access latency and bandwidth targets.
More importantly, \name{} is able to handle local memory contention as well as memory bandwidth interference at various multi-tenant settings. 
Extensive evaluations on 80 workloads across seven application categories show that \name{} achieves up to \bestimprove{} better application performance than TPP and 20.3\% better than Colloid, while satisfying more applications' SLOs.
\name{} also achieves 8.4$\times$ longer SLO satisfaction time than TPP in a long-running experiment to handle dynamic workload changes.

We make the following research contributions:
\begin{denseitemize}
    \item We conduct a thorough QoS analysis on production-ready tiered memory systems to qualitatively analyzing the performance impact of local memory contention and memory bandwidth interference.
    \item We propose a novel admission control and real-time adaptation algorithm tailored for tiered memory systems to achieve different SLOs for coexisting applications.
    \item We implement a new kernel-level resource management scheme to control resources on tiered memory. We plan to share our design with the Linux community for open discussion after the paper's publication.
\end{denseitemize}

\section{Background and Motivation}
\label{sec:motivation}

We start by introducing two sources of performance unpredictability in tiered memory (\S\ref{sec:two-sources}), followed by a quantitative QoS analysis using CXL hardware and real-world applications (\S\ref{sec:moti-exp}). 
We then motivate \name{}'s design by analyzing why existing solutions fail to provide QoS support (\S\ref{sec:compare-colloid}).

\subsection{Sources of Performance Unpredictability}
\label{sec:two-sources}
There are two primary sources of unpredictable performance -- local memory contention and memory bandwidth interference -- as applications share memory resources in tiered memory.

\para{Local memory contention.}
Multiple applications deployed on the same tiered memory system compete for pages on the fast tier (i.e., local memory).
The core design principle of tiered memory systems is to keep hot pages on local memory (i.e., DRAM on the fast tier) while migrating cold pages to the slower tier~\cite{hemem, nimble, autotiering, tpp, thermostat, HeteroOS, tmts}.
However, such a design does not distinguish among applications with different SLOs.
A low-priority application with a large amount of hot memory can still compete for local memory, hurting the performance of more important applications and causing priority inversion.

\para{Memory bandwidth interference.}
Applications consuming high memory bandwidth can affect the performance of coexisting applications.
The problem is further complicated in tiered memory systems, where applications can generate bandwidth with memory requests accessing different memory tiers.
We identify two types of memory bandwidth interference -- \textit{intra-tier interference} and \textit{inter-tier interference}.
Intra-tier interference refers to the one happening on the same tier, which is also common in conventional, single-tier systems~\cite{fairqueuing, mcp, heracles, parties}. 
Inter-tier interference occurs when excessive memory requests on one memory tier cause a slowdown of requests on another tier.
Existing solutions on mitigating memory bandwidth interference work either 
(1) on the memory request level, such as memory request prioritization~\cite{fairqueuing} or memory channel partitioning~\cite{mcp}; or 
(2) on the CPU core level to throttle memory bandwidth from best-effort tasks~\cite{parties, heracles}.
None of these solutions is designed to work across multiple memory tiers and thus cannot handle inter-tier interference. %

\begin{figure}[!t]
    \centering
    \subfloat[Impact on Latency Performance]{
        \label{fig:moti-lt-sweep}
        \includegraphics[width=3in]{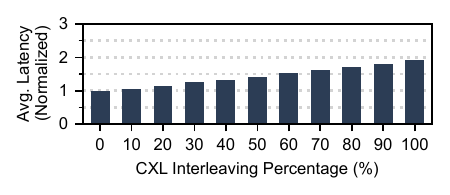}
    } \\
    \subfloat[Impact on Bandwidth Performance]{
        \label{fig:moti-bw-sweep}
        \includegraphics[width=3in]{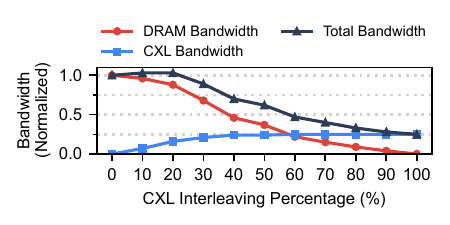}
    }
    \captionsetup{width=\columnwidth}
    \caption{Latency and bandwidth performance at different CXL interleaving ratios to illustrate the impact of local memory.}%
    \label{fig:moti-local-memory-impact}
\end{figure}

\subsection{Quantitative Analysis: QoS in Tiered Memory}
\label{sec:moti-exp}
We now quantitatively analyze how local memory contention and memory bandwidth interference affect application performance and draw key insights to design \name{}.
In this work, we classify applications into two types: (1) \textbf{Latency-sensitive (LS)} applications that desire low memory access latency, and (2) \textbf{bandwidth-intensive (BI)} applications that require sustained memory bandwidth.
Our experiments run on dual-socket AMD Genoa servers with 96 physical cores and 12 DDR5 memory channels per socket. 
CXL memory is enabled via two Astera Labs memory expansion cards with 2 DDR5 channels. 
Each card connects via x16 PCIe lanes (x32 lanes in total). 
The servers have a 1.8TB memory footprint with 768GB DDR5 per socket and 256GB CXL memory.

\subsubsection{Impact of Available Local Memory}
\label{sec:local-memory-impact}
We develop two microbenchmarks to represent LS and BI applications, which we denote as \textit{LS} and \textit{BI} in this section for brevity.
\textit{LS} performs random memory access among a 4GB memory region.
\textit{BI} allocates enough CPU cores to generate memory bandwidth at maximum capacity on 128MB region per core.
These microbenchmarks allow us to easily control the proportion of memory access on CXL memory (the rest goes to local memory), which we denote as the CXL interleaving percentage.

Figure~\ref{fig:moti-local-memory-impact} shows the impact of available local memory on latency and bandwidth performance by varying the CXL interleaving ratio of the two microbenchmarks.
We observe the memory access latency of \textit{LS} is proportional to its available local memory, and it becomes 2$\times$ slower when all memory is moved to the CXL tier in the worst case.
On the other hand, \textit{BI}'s bandwidth performance degrades as more memory is accessed from CXL, dropping to 25\% of its original performance when all memory is accessed via CXL.
\begin{tcolorbox}[width=\linewidth, boxsep=0pt, left=5pt, right=5pt, sharp corners=all, colback=white!92!black, boxrule=1pt, frame empty]
\textbf{Takeaway \#1}: Local memory directly affects the performance of both types of applications and should be allocated judiciously during memory contention.
\end{tcolorbox}

\subsubsection{Deep Dive in Memory Interference}
\label{sec:inter-tier-interference}

\begin{figure}[!t]
    \centering
    \includegraphics[width=3in]{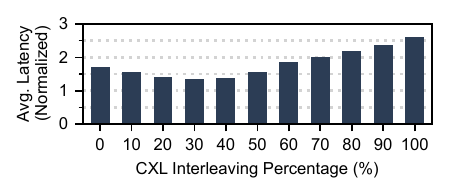}%
    \captionsetup{width=\columnwidth}
    \caption{Performance of \textit{LS} when \textit{BI} requires bandwidth at different CXL interleaving percentage. 
        Migrating \textit{BI} to CXL does not always lead to better performance of \textit{LS} due to inter-tier interference. 
        \textit{BI}'s performance is very close to Figure~\ref{fig:moti-bw-sweep} and omitted for brevity.}%
    \label{fig:moti-lt-local-bw-sweep}
\end{figure}

We now take a closer look at how LS applications are affected by memory bandwidth interference from BI applications.

\para{Varying \textit{BI} across tiers.}
We first configure \textit{LS} to always access local memory and vary \textit{BI}'s CXL-interleaving percentage.
The results are shown in Figure~\ref{fig:moti-lt-local-bw-sweep}, and we make two key observations.
First, initially the performance of LS improves due to the extra memory bandwidth from CXL channels. However, after a certain point, pushing more to the CXL degrades LS performance.
Second, the interference becomes the worst when all memory bandwidth demand is generated from CXL; the total memory bandwidth is dictated by the CXL channel's bandwidth capacity. 

\begin{figure}[!t]
    \centering
    \includegraphics[width=3.3in]{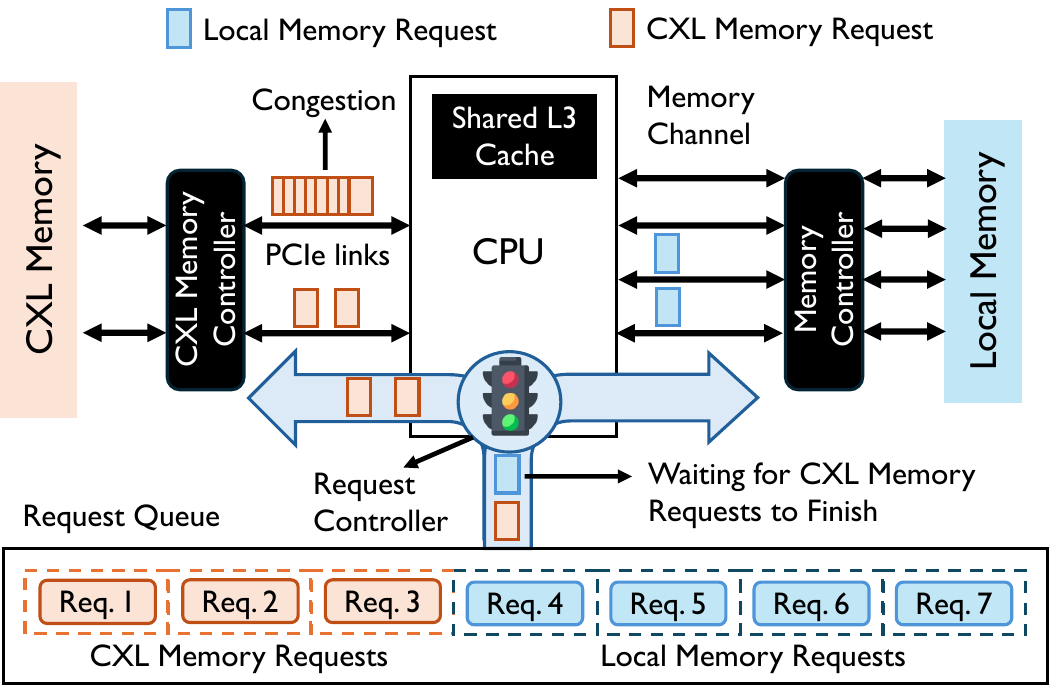}%
    \captionsetup{width=\columnwidth}
    \caption{Architectural diagram of how memory requests are handled in CXL-enabled tiered memory.}%
    \label{fig:cxl-arch}
\end{figure}

This behavior can be explained by examining how memory requests are handled in a tiered memory system. 
Figure~\ref{fig:cxl-arch} illustrates the high-level overview of a CXL-enabled tiered memory architecture.
There exist separate fixed-size queues in the hardware for memory requests to/from local memory and CXL memory. 
The effective bandwidth of the system depends on the depth of these queues.
The interleaving policy \cite{linux-interleaving} of OS processes memory requests by serving \textit{m} local memory requests, followed by \textit{n} CXL memory requests, where \textit{m:n} represents the interleaving ratio between local and CXL memory (by default, it is 1:1). 
The high bandwidth demand on CXL memory can lead to a longer queuing delay when there is a shorter queue size.
In addition, the total PCIe link bandwidth may not match the effective bandwidth of local memory channels and it will take comparatively longer to process a request.
This delay, in turn, impacts the scheduling of local memory requests, resulting in inter-tier interference.

Going back to Figure~\ref{fig:moti-lt-local-bw-sweep}, initially, when bandwidth demand on the local memory is high, its corresponding queue fills up, causing interference on local memory (i.e., intra-tier interference) to dominate.
When \textit{BI} starts to move requests from local memory to CXL, intra-tier interference decreases, causing latency to drop.
When more requests are moved to the CXL side, the queue holding CXL requests builds up.
Since both types of requests are handled by the same set of CPU cores, busy processing of the CXL requests can slow down the concurrent requests for local memory. 
\begin{tcolorbox}[width=\linewidth, boxsep=0pt, left=5pt, right=5pt, sharp corners=all, colback=white!92!black, boxrule=1pt, frame empty]
\textbf{Takeaway \#2}: We should determine the right amount of memory to migrate BI applications that \textit{reduces intra-tier interference while keeping inter-tier interference low.}
\end{tcolorbox}

\begin{figure}[!t]
    \centering
    \includegraphics[width=3in]{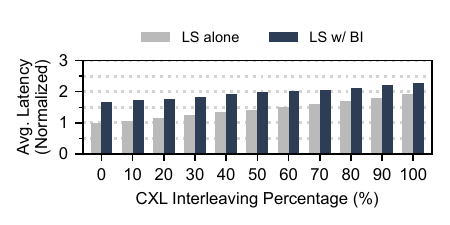}%
    \captionsetup{width=\columnwidth}
    \caption{Performance of \textit{LS} at different CXL interleaving ratios when \textit{BI} is fixed on local memory. Migrating more requests away from local memory does not improve performance as more requests are accessing the slower tier.}
    \label{fig:moti_bw_local_lat_sweep}
\end{figure}

\para{Varying \textit{LS} across tiers.}
Figure~\ref{fig:moti_bw_local_lat_sweep} shows the results of interference in a different setting, where we keep \textit{BI} fixed on local memory and vary \textit{LS}'s CXL interleaving percentage. %
This is used to simulate the scenario where one attempts to mitigate the interference on the fast tier by migrating \textit{LS}'s requests away from local memory to CXL memory.
However, such an approach leads to worse performance because more memory requests are now accessing the slow tier. %
\begin{tcolorbox}[width=\linewidth, boxsep=0pt, left=5pt, right=5pt, sharp corners=all, colback=white!92!black, boxrule=1pt, frame empty]
\textbf{Takeaway \#3}: We should proactively mitigate the interference on local memory to provide predictable performance for LS applications.
\end{tcolorbox}

\begin{figure}
    \centering
    \includegraphics[width=2.6in] {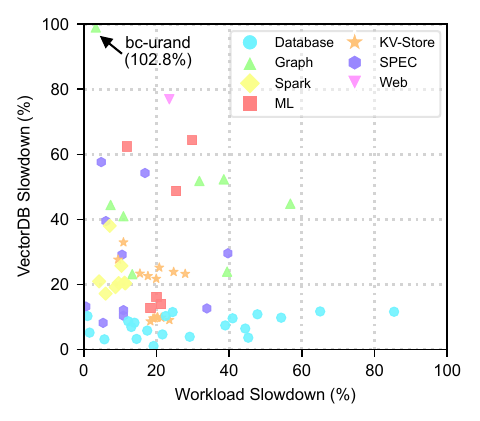}
    \captionsetup{width=\columnwidth}
    \caption{Unpredictable performance of 80 workloads and VectorDB as they compete for local memory on the fast tier. Existing solutions cannot distinguish among applications when migrating their hot pages, and thus cannot provide QoS guarantees.}
    \label{fig:moti-local-dram-contention}
\end{figure}

\subsubsection{Impact on Real Applications}
\label{sec:moti-real-app-exp}
To illustrate the impact of memory capacity and bandwidth contentions on real-world applications in a tiered memory system, we perform the same set of experiments on local memory contention and memory bandwidth interference across 80 applications across seven diverse categories.
Appendix~\ref{app:real-apps} describes the full list of applications used. 
We run these experiments on dual-socket Intel Xeon Gold 6330 servers with 56 physical CPU cores where each socket contains 512GB DDR4 memory. 
For controlling the environmental setup, we use the remote socket as the slow tier, with uncore frequency tuned down to match the memory access latency in our real CXL hardware.
All the experiments here allocate different physical cores to foreground and background workloads, ensuring that no core contention occurs.

Figure~\ref{fig:moti-local-dram-contention} presents the impact of local memory contention. 
In this experiment, VectorDB is selected as the background application, and each time another application is launched as the foreground workload.
We enable TPP~\cite{tpp}, a state-of-the-art tiered memory system, to perform page migration.
The total working set size (WSS) of the applications in each setting exceeds the available local memory on the fast tier.
We observe the performance degradation between the foreground and background applications highly depends on their memory access frequency.
The foreground application with lower memory access frequency than VectorDB (e.g., TPC-H) always experiences larger performance degradation, and vice versa (e.g., SPEC, DLRM).
In cases of similar memory access frequencies, both applications are affected (e.g., Graph).
The results are unsurprising as TPP does not distinguish between applications with different SLOs during page migration.

\begin{figure}
    \centering
    \subfloat[Intra-tier Interference]{
            \label{fig:moti-intra-inter-interference}
            \includegraphics[width=3.3in]{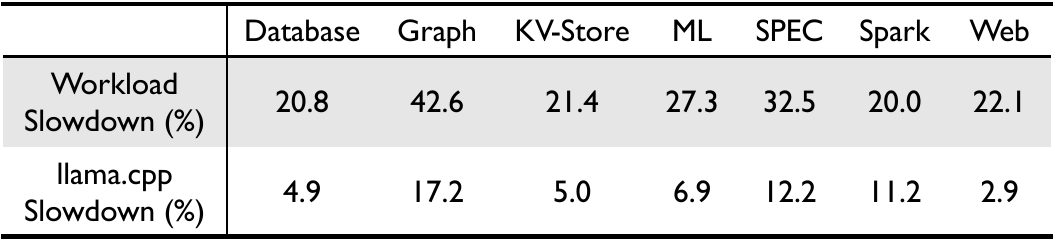}%
        }\\ %
    \subfloat[Inter-tier Interference]{
        \label{fig:moti-inter-tier-interference}
        \includegraphics[width=3.3in]{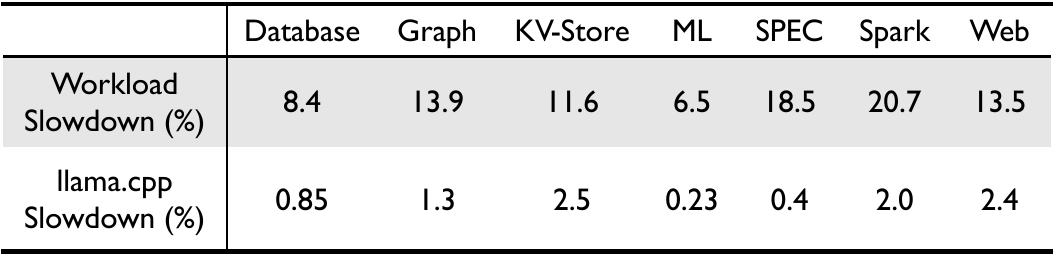}%
    }
    \captionsetup{width=\columnwidth}
    \caption{\llama{}'s memory bandwidth creates interference, resulting in a significant drop in the throughput performance of all kinds of workloads. (a) shows intra-tier interference when \llama{} is on the same fast tier with different workloads. (b) shows inter-tier interference after all \llama{}'s memory is migrated to CXL.}%
    \label{fig:moti-memory-interference}
\end{figure}

Memory bandwidth interference can also greatly impact coexisting applications.
Figure~\ref{fig:moti-memory-interference} illustrates the impact of both types of memory bandwidth interference.
We select \llama{}~\cite{llama} inference tasks (running on local DRAM) as the background task to create high memory bandwidth demand.
Under TPP, all foreground applications we tested experienced up to 50\% performance degradation due to intra-tier interference.
When all \llama{}'s memory is migrated to the slow tier, up to 32\% slowdown is observed due to inter-tier interference.

\begin{figure}[!t]
    \captionsetup{width=\columnwidth}
    
    \includegraphics[width=3.3in]{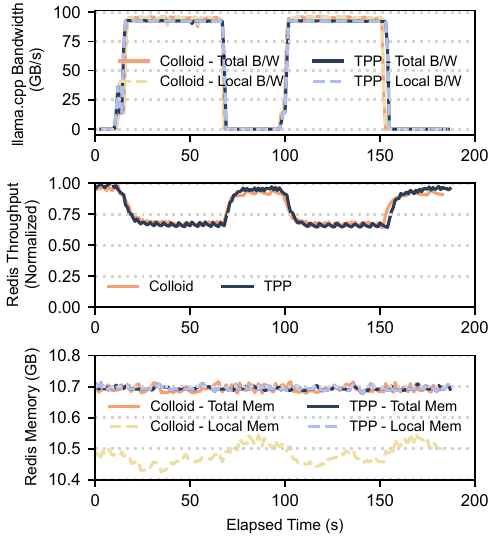}%
    \caption{\llama{}'s memory bandwidth creates interference, resulting in a significant drop in Redis throughput. }    
    \label{fig:moti-colloid-memory-interference}
\end{figure}

\subsection{Lack of QoS Support in Existing Solutions}
\label{sec:compare-colloid}

We now summarize why existing tiered memory systems fail in providing QoS, and thus calling for a new QoS-aware tiered memory system.
There are two major design principles adopted by existing solutions~\cite{hemem, nimble, autotiering, tpp, thermostat, HeteroOS, colloid, tmts} that are proved to be beneficial for memory-intensive workloads: 
(1) optimizing \textit{system-level} performance such as overall memory access latency and bandwidth; and 
(2) \textit{reactively} allocating resources based on memory-level measurements.
The first principle assumes a single application to be optimized, and the second principle assumes workloads remains in steady state after their ramp-up period. 
However, neither assumptions hold when it comes to providing QoS support.

We conduct a motivating experiment illustrating how TPP and Colloid~\cite{colloid}, two state-of-the-art solutions, are inefficient in providing predictable performance.
In this experiment, we measure how the performance of Redis is affected by two consecutive inference requests from \llama{}, as shown in Figure~\ref{fig:moti-colloid-memory-interference}.
In this setup, Redis is considered more important than \llama{}, which is configured to be an offline batch inference workload. 
Each of the two requests from \llama{} caused 100 GB/s bandwidth spikes, which leads to two drops in Redis throughput.

Neither TPP nor Colloid identifies Redis as an application that is under bandwidth interference.
TPP does not migrate any pages of \llama{} and Redis because WSS of both applications fit in local memory, and TPP is not designed to handle interference. 
On the other hand, Colloid migrates the pages of Redis to the slow tier to balance out the memory access latencies caused by increased contention due to \llama's requests.
In addition, the demotion of pages are too slow to make noticeable impact to recover application performance. 

In the next section, we discuss the decision choices we make in designing \name{} to enable \textit{QoS-awareness} and \textit{proactive resource allocation}.

\section{\name{} Overview}
\label{sec:overview}
With multiple applications sharing a tiered memory system, our goal in this work is to provide predictable performance among applications with different SLOs under memory resource contention and memory interference.

\subsection{Design Principles}
\label{sec:design-principle}
\para{QoS granularity.}
The first step to QoS-awareness is determining at what granularity we should support QoS and manage memory resources.
There exist multiple levels where QoS can be enforced, such as (1) an entire application, (2) individual data structures, (3) individual \texttt{mmap()} calls, or even (4) individual memory accesses.
Although finer granularity allows more precise QoS assignments across different segments of the application, we build \name{} at the \textit{application level} to ensure (1) easy deployment without modifying applications, (2) a clean QoS interface, and (3) low overhead in managing memory resources and performance monitoring.
Additionally, providing QoS at the application level allows us to reuse existing kernel mechanisms for identifying and migrating hot and cold pages, which has proved to be efficient~\cite{tpp}.

\para{Proactive memory resource allocation.}
Instead of overprovisioning resources to handle worst-case overload scenarios, we decide to proactively allocate \textit{the right amount of resource that can satisfy an application's SLO}.
This allows \name{} to accommodate more applications while meeting their SLOs.
Following this design principle, \name{} provides QoS via a combination of \textit{admission control} and \textit{real-time adaptation}.
The former admits applications using the right amount of resources, and the latter adjusts resource allocations for applications when runtime workload changes affect QoS.

\para{Choice of performance indicators.}
Selecting the right performance indicators is critical for providing QoS, as \name{} needs to react quickly to meet applications' SLO.
In \name{}, we prefer \textit{low-level metrics} as performance indicators compared to application-level metrics.
Low-level metrics can be collected via hardware-based PMU counters. %
Their measurements require no application modification, and as we will soon show, they closely reflect the application performance and can react faster to real-time workload changes (\S\ref{sec:evaluation}).
In particular, we collect \textit{memory access latency} and \textit{memory bandwidth} per application. %
On AMD processors, we measure memory access latency using memory load events from L3 cache misses provided by Instruction Based Sampling (IBS)~\cite{ibs}, and memory bandwidth is measured by $\upmu$Prof~\cite{uprof}. 
On Intel platforms, latencies can be measured through processor event-based sampling (PEBS)~\cite{pebs} where bandwidth can be measured via Intel Platform QoS (PQoS)~\cite{pqos}.

\para{Prioritization.}
When multiple memory-intensive applications coexist, there is no guarantee that we will always have enough resources to satisfy everyone's SLO.
For example, an important application may arrive later than less critical ones, and an application may require more resources to fulfill increasing load. 
To this end, \name{} leverages \textit{strict priority} to ensure high-priority applications get guaranteed performance even when the tiered memory runs out of resources.\footnote{Lower priority applications that yield (part of) their resources to higher priority ones continue to run as best-effort to preserve work conservation.}
We choose to apply a separate priority scheme on top of SLOs because more stringent SLOs do not always mean higher priority, and BI applications are not necessarily less important than LS ones.
In \name{}, priority levels are uniquely assigned to avoid ties among applications.

\begin{figure}[tb]
    \centering
    \includegraphics[width=0.98\columnwidth]{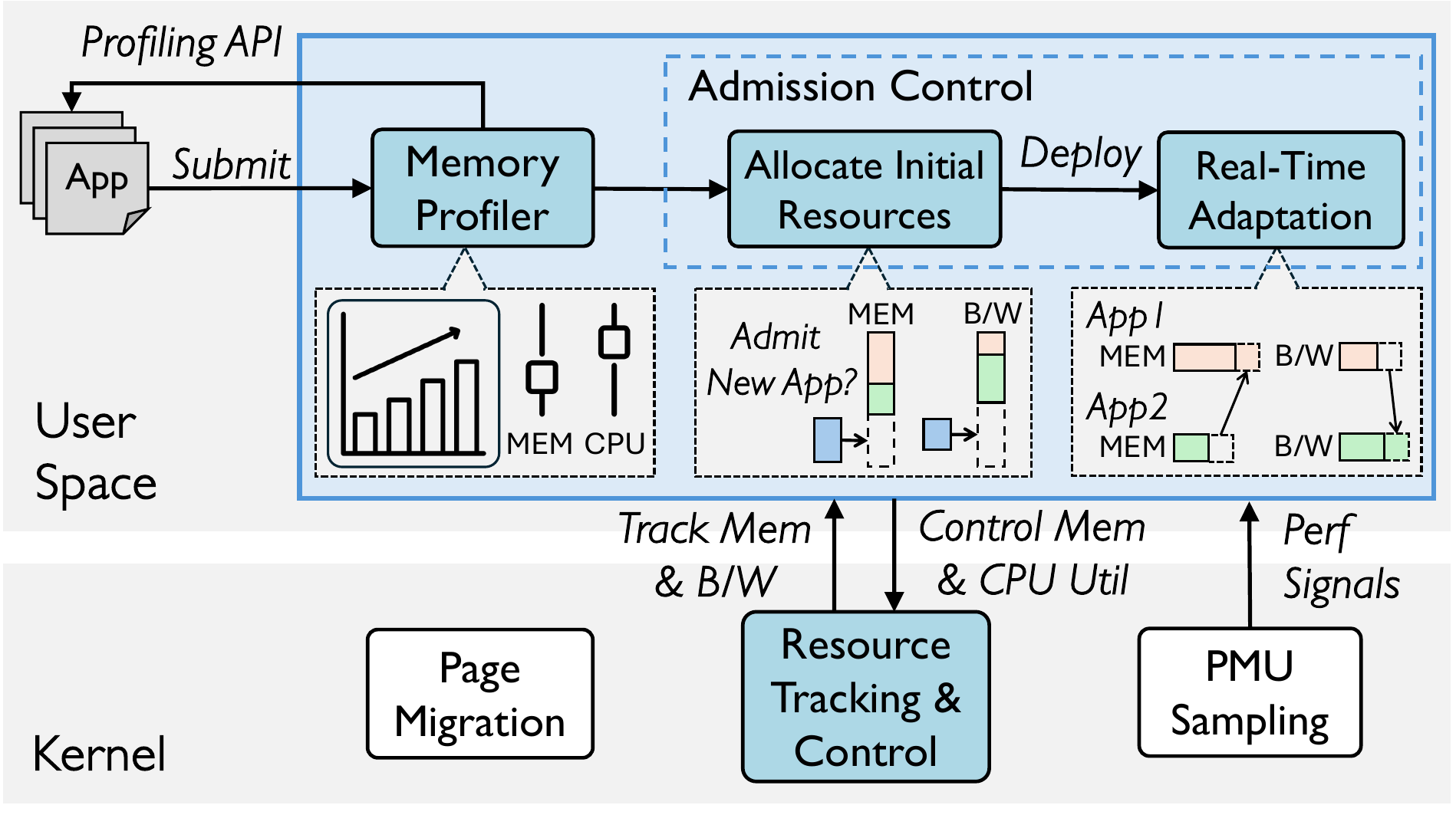}%
    \captionsetup{width=\columnwidth}
    \caption{High level system overview of \name{}. The memory profiler determines the memory resources needed to maintain SLOs when applications are running in isolation. The admission controller takes this information and proactively determines the right resource to allocate for applications while mitigating interference. It also performs real-time adaptation to adjust resource allocation during runtime. The resource controller tracks and controls resources at the application level.}%
    \label{fig:system-overview}
\end{figure}

\subsection{System Overview}

Figure~\ref{fig:system-overview} presents a system diagram of \name{}.
Its design includes a memory profiler, an admission control algorithm, as well as a memory resource controller implemented inside the Linux kernel.

A \name{} user (e.g., a cluster operator) submits an application along with a list of information, including the number of CPU cores, memory requirement, application type, its priority level, and the SLO.
The SLO of LS applications specifies a \textbf{memory access latency} target, and that of BI ones is a \textbf{maximum memory bandwidth} the application needs.

\name{} proactively controls two types of resources, \textbf{available local memory} and \textbf{cpu utilization} at the application level. %
Since the existing kernel mechanism on memory usage tracking and controlling does not distinguish among different tiers, \name{} enables per-tier page reclamation to control an application's available local memory (\S\ref{sec:resource-management}).
\name{} adopts existing Linux page LRUs and NUMA Balancing~\cite{numa-balancing} for page temperature detection and migration. %
When \name{} receives an application, it first profiles offline to find the minimum amount of memory resource needed to satisfy the SLO when running in isolation (\S\ref{sec:profiler}). 
The profiling procedure only takes less than 1\% of the workload's entire lifetime, incurring negligible overhead.
The profiler also provides users with an API to measure the two SLO metrics to help them set an appropriate SLO. 
After profiling, \name{}'s admission control determines whether the new application should be admitted onto the node, and if so, how much resource we should allocate to the application, along with any changes to the existing resource allocation of other applications based on the application's priority level (\S\ref{sec:allocate-initial-resources}).
The admission control keeps monitoring an application's performance during its lifetime and performs real-time adaptation on its resource allocation to ensure SLO is maintained in case of workload changes and bandwidth interference (\S\ref{sec:real-time-adaptation}).

\section{\name{} Design}
\label{sec:design}

In this section, we describe \name{} in details, including its application-level resource management, memory profiling, and admission control with a real-time adaptation module.

\subsection{Application-Level Resource Management}
\label{sec:resource-management}
\name{} manages two resources -- local memory and CPU utilization -- to enforce SLO.
The available local memory for an application determines the ratio of memory accesses on local vs. CXL memory, which directly affects an LS application's performance.
As local memory is limited and shared among all applications, \name{} decides for each LS application a \textit{local memory limit} based on their SLO.
Limiting an application's local memory can also affect its bandwidth, as accessing more memory from a slower tier decreases the aggregate bandwidth (\textbf{Takeaway \#1} in \S\ref{sec:local-memory-impact}).
Meanwhile, it saves additional local memory to accommodate other applications.
However, this approach alone is not enough to control an application's bandwidth for two reasons.
First, placing too many memory requests on CXL memory creates inter-tier memory interference (\S\ref{sec:inter-tier-interference}).
Second, the bandwidth cannot be tuned further down after migrating all pages to CXL memory (Figure~\ref{fig:moti-bw-sweep}).
Therefore, \name{} also limits the CPU utilization of BI applications in addition to local memory limit in order to achieve a finer-grained control over bandwidth.

After deciding what resources \name{} manages for applications, the next step is to seek an efficient way of tracking and controlling those resources.
Linux kernel already provides cgroup~\cite{cgroup} to track and limit the memory and CPU usage of a collection of processes.
However, the memory control mechanism in cgroup cannot be directly applied in \name{}, because it does not distinguish between different memory tiers when tracking and controlling memory usage.
Instead, one can only specify a total memory limit that accounts for the total memory usage across all tiers for a given process, then the virtual pages will be swapped to the disk if the memory limit is reached.

\para{Per-tier memory management in Linux cgroup.}
We make a series of modifications on cgroup to enable per-tier memory tracking and control.
While observing the total memory usage of an application, we break it down into individual tier-level usage. 
In a CXL-enabled tiered memory system, memory nodes appear to the system as separate NUMA nodes.
We can thus assume that each tier consists of one or multiple NUMA nodes.
We enhance cgroup to track the list of pages within each LRU associated with the NUMA nodes across a specific memory tier.
Specifically, we introduce a new memory limit-controlling interface, \texttt{memory.per\_numa\_high} that controls the max memory usage for an application on each NUMA node. 
Note that \texttt{memory.per\_numa\_high} works concurrently with the global cgroup interface (\texttt{memory.high}). 
While the latter controls the total system-wide memory usage of an application, our interface restricts the contribution of an application's memory footprint on each memory tier.

When allocating pages, \name{} uses Linux's default ``allocate on fast memory tier first" page allocation policy unless specified otherwise. 
However, if the memory usage of a NUMA node within a specific memory tier exceeds its \texttt{memory.per\_numa\_high} threshold, \name{} stops memory allocation and initiates page reclamation only on that NUMA node. 
Here, the default reclamation mechanism is to demote (i.e., page migration) to the next available memory tier.
Moreover, a change to \texttt{memory.per\_numa\_high} (e.g., by \name{} or system admins) immediately triggers reclamation across all the nodes where the new limit is below the current memory usage.
Similar to TPP~\cite{tpp}, we leverage NUMA Balancing~\cite{numa-balancing} to allow page promotion among different nodes. 

\name{} controls an application's CPU utilization using the native cgroup \texttt{cpu.max} interface. 
By configuring this parameter, \name{} can adjust the application's utilization across all CPU cores it employs. Compared with vendor-provided bandwidth control mechanisms -- such as Intel's Memory Bandwidth Monitoring (MBM) \cite{mba-intel} -- managing CPU utilization in this manner offers a more fine-grained and predictable approach \cite{mt2}.
Details of how \name{} selects an application's local memory limit and CPU utilization to provide QoS will be described next.

\subsection{Memory Profiler}
\label{sec:profiler}
Before deploying an application, we first need to determine the right amount of memory resource needed to meet its SLO to provide QoS for more applications.
In \name{}, this task is handled by the memory profiler.

Taking a user application and its SLO as inputs, the profiler finds the right amount of local memory the application needs to satisfy its SLO when running in isolation.
For both types of applications, it starts by putting all their memory on the fast tier with full CPU utilization.
If the SLO is not met even at this initial stage, the application is marked as \emph{inadmissible}; the user should adjust the SLO or deploy it on another machine with larger memory or bandwidth.
Otherwise, the profiler decreases its local memory limit until the measured performance matches the SLO.
For BI applications, if the actual bandwidth is still above the SLO after the local memory limit drops to zero, the profiler starts to decrease CPU utilization until the SLO is met.

Note that the local memory limit found during profiling represents the minimum memory the application needs in isolation and is used by the admission control (\S\ref{sec:admission-control}) to determine the actual local memory and CPU utilization to allocate during deployment.
\name{} also records the profiled memory bandwidth for BI applications, which represents the target bandwidth to meet its initial load during admission. 

Finally, \name{} also performs a one-time profiling per machine to characterize inter-tier memory bandwidth interference, which will later be used by our admission control and real-time adaptation.
Specifically, \name{} determines \textit{two thresholds}, (i) a threshold on \textit{healthy local bandwidth} ($Thresh_{local\_bw}$) and (ii) a threshold on \textit{the rate of remote NUMA hint fault} ($Thresh_{numa}$), to monitor whether the system has reached the boundary of triggering severe intra-tier and inter-tier interference.
The profiler leverages the \textit{LS} and \textit{BI} microbenchmarks (\S\ref{sec:moti-exp}) to perform this task.
To determine $Thresh_{local\_bw}$, the profiler launches both \textit{LS} and \textit{BI} on the fast tier, and increases \textit{BI}'s bandwidth until a noticeable interference on \textit{LS}'s latency is found.
Similarly, $Thresh_{numa}$ is determined by sweeping the CXL interleaving percentage of \textit{BI} until an obvious performance degradation is observed on \textit{LS} running on the fast tier (10\% performance degradation is set for both thresholds in our implementation).

\begin{lstlisting}[language=Python, commentstyle=\scriptsize\itshape\color{gray}, caption={Admitting LS and BI applications.}, label={lst:admission_control}]
if Local_avail_mem > app.profiled_mem_limit:
    Allocated_mem = app.profiled_mem_limit
else:
    priority_queue.(*@\textcolor{functioncolor}{yieldMem}@*)(app.profiled_mem_limit)
    Allocated_mem = Current_local_avail_mem
    
if app.type == "BI":
    # Local_avail_bw considers intra-tier intf.
    if Local_avail_bw > app.profiled_bw:
        Allocated_bw = app.profiled_bw
    else:
        # yieldBW considers inter-tier intf.
        priority_queue.(*@\textcolor{functioncolor}{yieldBW}@*)(app.profiled_bw)
        Allocated_bw = Current_local_avail_bw
\end{lstlisting}

\subsection{Admission Control}
\label{sec:admission-control}

At the core of \name{}'s design is its admission control algorithm, which proactively assigns resources among multiple applications at the presence of local memory contention and memory bandwidth interference.
A naive solution is to take each incoming application's profiling result and keep deploying applications until local memory or bandwidth capacity runs out; i.e., {first-come-first-service (FCFS)}.
However, this approach has two drawbacks.
First, a critical application may arrive late, and the resources already assigned to less critical applications cannot be taken back.
Second, it does not consider the impact of bandwidth interference, which can lead to SLO violation.

In \name{}, we design a new admission control algorithm tailored for tiered memory.
It leverages \textit{strict priority scheme} to prioritize important applications, and \textit{maximizes local memory utilization} to admit more applications with SLO guarantees while \textit{avoiding intra- and inter-tier interference}. 
To achieve the aforementioned goal, there are two key questions we need to answer during admission control: 
(1) How to allocate initial resources when admitting applications to shared tiered memory? 
(2) How to ensure admitted applications can preserve their SLO guarantees in the presence of memory interference and workload changes?

\subsubsection{Determining Initial Resource Allocation}
\label{sec:allocate-initial-resources}
Listing~\ref{lst:admission_control} presents a high-level overview of how \name{} determines the initial resources when admitting applications.
The key idea in this step is to \textit{proactively} allocate initial resources to satisfy as many applications' SLOs as possible while mitigating potential memory interference.
At the arrival of a new application, \name{} immediately drops it if it has been labeled inadmissible by the profiler since its SLO (either latency or bandwidth) can never be met. %
Otherwise, \name{} starts to assign resources based on the application type.

\para{Admitting LS applications.}
\name{} directly admits an LS application if there is available local memory to satisfy its profiled local memory requirement.
Otherwise, if the current local memory is smaller than its profiled memory limit, \name{} tries to find memory from applications with a lower priority to \textit{yield local memory}, starting from the one with the lowest priority to give this memory. 
If the victim application's local memory limit drops to zero, \name{} moves on to the next victim until no existing applications with lower priority can yield local memory (line 1-5 in Listing \ref{lst:admission_control}). %

\para{Admitting BI applications.}
Given a BI application, the goal is to assign the right amount of local memory and CPU utilization to satisfy its bandwidth SLO.
\name{} starts by assigning local memory to BI applications in the same way as it does to LS ones. 
However, assigning the profiled local memory to the incoming application does not always mean it can achieve its bandwidth SLO as existing applications may have already taken a large portion of the total bandwidth capacity.
Under this scenario, \name{} finds existing BI applications with lower priority in ascending order to \textit{yield bandwidth}.
Similar to the procedure in \textit{yield local memory}, if the victim BI application's bandwidth cannot be reduced anymore, it will move on to the next application until no existing applications with lower priority can yield bandwidth (line 7-14 in Listing \ref{lst:admission_control}). 
This is done by decreasing the local memory limit of the victim application(s).
During this process, \name{} monitors the current rate of remote NUMA hint faults and avoids inter-tier interference by switching to decreasing the victim application's CPU utilization once $thresh_{numa}$ is exceeded (\textbf{Takeaway \#2} in \S\ref{sec:inter-tier-interference}).
In the meantime, \name{} watches out for potential \textit{intra-tier interference} on the fast tier by verifying if (1) there exist LS applications with higher priority, and (2) the healthy bandwidth threshold ($thresh_{local\_bw}$, \S\ref{sec:profiler}) for the fast tier is exceeded.
If both conditions are met, \name{} stops assigning bandwidth on the fast tier to the incoming application. 
The additional checks in \textit{yield memory} ensures newly admitted applications do not start by creating severe interference, before more fine-grained control on resources kicks in to maintain application SLOs, which we will describe next.

\subsubsection{Adaptating to Interference and Workload Changes}
\label{sec:real-time-adaptation}

So far, we have discussed how \name{} admits applications and by proactively allocating resources during application launch time to prevent undesired interference.
However, merely relying on this initial resource allocation does not always preserve SLOs. 
SLO violations can still happen in the following two scenarios.
First, LS applications can still miss SLOs due to bandwidth interference when BI applications' bandwidth sits around one of the two interference thresholds.
Second, the amount of resource needed to maintain an application's SLO can shift over time due to workload changes after launch.
Therefore, \name{} should adjust resource allocations dynamically in real time to ensure QoS guarantees for critical applications in such scenarios.

\begin{lstlisting}[language=Python,  commentstyle=\scriptsize\itshape\color{gray}, caption={Real-Time Adaptation in Admission Control}, label={lst:real_time_adaptation}]
for app in priority_queue:
    if app.SLOSatisfied:
        (*@\textcolor{functioncolor}{yieldResource}@*)(app)
    elif !app.SLOSatisfied:
        if app.type == "BI" and app.cpu_util < 1:
            app.(*@\textcolor{functioncolor}{increaseCPUUtil}@*)()
        elif (*@\textcolor{functioncolor}{BWReducible}@*)():
            priority_queue.(*@\textcolor{functioncolor}{yieldBW}@*)()
        elif app.total_mem < app.local_mem_limit:
            priority_queue.(*@\textcolor{functioncolor}{yieldMem}@*)()
    if all_apps_satisfied and (*@\textcolor{functioncolor}{lowestPrio}@*)(app):
        priority_queue.(*@\textcolor{functioncolor}{workConservation}@*)()
\end{lstlisting}

\name{} evaluates all applications and makes necessary real-time adjustments to resource allocation every 200 ms.
Listing~\ref{lst:real_time_adaptation} presents an overview of this process.
Applications are processed in the order of descending priority to ensure critical applications are protected first.
For each application, \name{} verifies whether the measured performance (i.e., memory access latency for LS applications and memory bandwidth for BI applications) meets its SLO. 
If the measured performance of an application is better than its SLO, \name{} attempts to yield its resources to benefit existing or future applications (line 3).
\name{} \textit{monitors inter-tier interference} during this process (using $thresh_{numa}$) to prevent hurting other applications.
For BI applications, once the local memory limit reaches zero, \name{} starts reducing CPU utilization to yield more bandwidth.

It is more challenging when the measured performance is worse than SLO, as there may be multiple possible causes.
\name{} takes a series actions to identify the correct cause and make necessary adjustments. %

(1) \textit{Increase CPU utilization.}
For a BI application, \name{} first attempts to increase its CPU utilization (line 6).
By increasing CPU utilization first, we improve the application's bandwidth without unnecessarily consuming additional local memory resource.

(2) \textit{Mitigate bandwidth interference.}
Next, \name{} decreases the bandwidth of low-priority applications until no lower-priority BI applications are available for reduction (lines 7–8). 
The bandwidth reduction process follows the same approach used in yielding bandwidth during initial resource allocation (\S\ref{sec:allocate-initial-resources}): it reduces the bandwidth of BI applications in order of priority by lowering their local memory limits, while simultaneously monitoring inter-tier interference to initiate CPU utilization reduction.
This step verifies if the performance drop is caused by interference.
Although increasing local memory limit can also improve performance, we have shown it is more effective to reduce bandwidth interference first (\textbf{Takeaway \#3} in \S\ref{sec:inter-tier-interference}).
Since all applications go through our real-time adjustment in strict priority order, \name{} ensures the BI application that generates excessive bandwidth will eventually be punished, and other BI applications will retain SLO if there is enough bandwidth.

(3) \textit{Increase local memory allocation.}
If the measured performance is still worse than SLO, we are certain the target application needs more memory due to a workload change. %
In this case, \name{} allocates memory to the application in the same way, it finds the local memory from low-priority applications in ascending order of the priority level.
Similar to how initial local memory is allocated (\S\ref{sec:allocate-initial-resources}), \name{} finds the local memory from low-priority applications in ascending order of the priority level.

After going through all the applications in the current adaptation period, if there is additional available memory and all applications' SLOs are satisfied, \name{} allocates the remaining memory to applications based on descending priority for work conservation.

\begin{figure}
    \centering
    \subfloat[LS3's SLO is unmet, requiring resource reallocation for the applications.]{
            \label{fig:real-time-example-event-1}
            \includegraphics[width=3.3in]{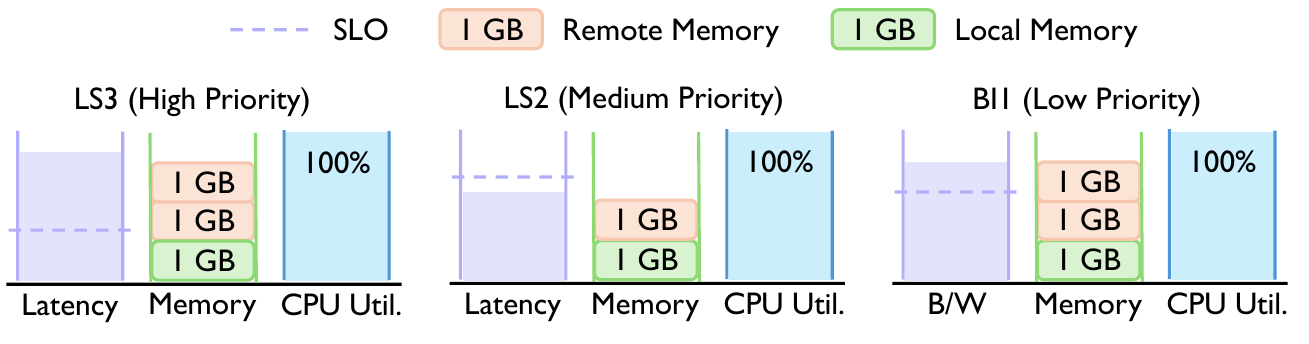}%
        }\\[0.5em]
    \subfloat[Reduce bandwidth for BI1 by demoting its local memory and lowering CPU utilization to avoid inter-tier interference. Although the latency of LS3 is reduced, it still does not meet the SLO, requiring further resource allocation.]{
            \label{fig:real-time-example-event-2}
            \includegraphics[width=3.3in]{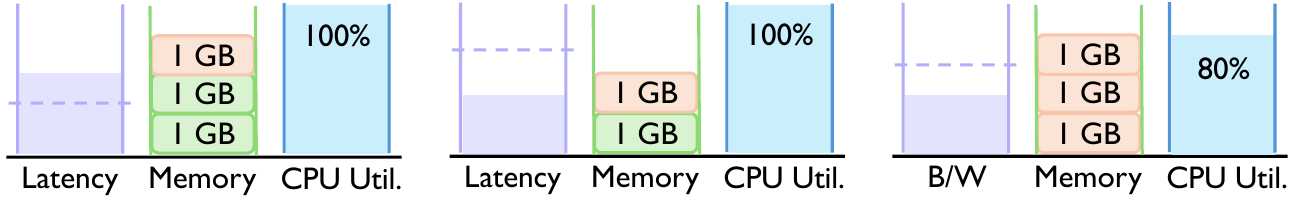}%
        }\\[0.5em]
     \subfloat[Allocate more memory to LS3 by yielding memory based on priority, with only LS2's memory available. After the demotion, LS3 meets its SLO.]{
            \label{fig:real-time-example-event-3}
            \includegraphics[width=3.3in]{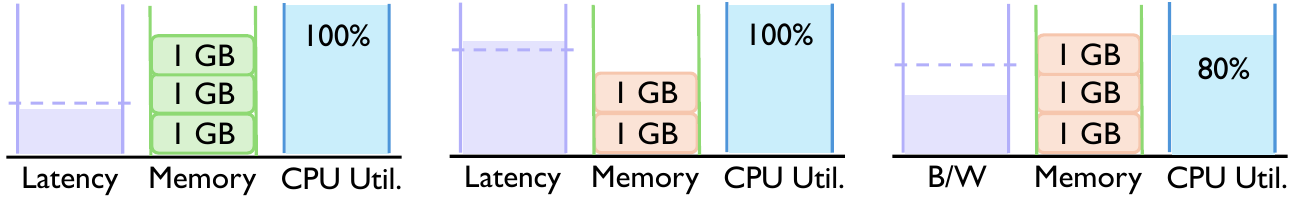}%
        }
    \captionsetup{width=\columnwidth}
    \caption{Examples of real-time adaptation. There are three applications in the system, with different SLO targets and priorities.}%
    \label{fig:real-time-example}
\end{figure}

\para{An example.}
Figure~\ref{fig:real-time-example} illustrates an example of real-time adaptation by \name. 
The system hosts three applications: BI1, LS2, and LS3, with priorities in ascending order (BI1 has the lowest priority, LS3 the highest). 
At a specific time (Figure \ref{fig:real-time-example-event-1}), \name{} detects that the latency of LS3 exceeds its SLO target and decides to allocate more resources to improve its performance.
First, \name{} reduces the bandwidth allocated to lower-priority BI applications in descending priority order.
As a result, BI1's memory is demoted to the CXL side. 
This demotion eventually exceeds $thresh_{numa}$, and \name{} decides to reduce BI1's CPU utilization to mitigate inter-tier interference (Figure \ref{fig:real-time-example-event-2}). 
The reduction in BI1's bandwidth improves the latencies of both LS2 and LS3. 
However, LS3's latency remains worse than its SLO, and thus \name{} proceeds to allocate additional memory to LS3.
It reclaims memory from lower-priority applications based on their priority order. 
At this point, among applications with lower priority than LS3, only LS2 has local memory available. 
Therefore, \name{} demotes 1\,GB of LS2's memory, reallocating it to LS3. 
Consequently, LS3's latency meets its SLO target (Figure \ref{fig:real-time-example-event-3}).

\section{Evaluation}
\label{sec:evaluation}
We evaluate \name{}'s capability of providing QoS among applications sharing a tiered memory system, in the presence of local memory contention, bandwidth interference, and dynamic workload changes.
Our key findings are as follows: 

\begin{enumerate}[leftmargin=5ex, nosep, label=(\arabic*)]
    \item \name{} closely tracks SLOs for both LS and BI applications by allocating right amount of resources (\S\ref{sec:eval-slo-compliance}).
    \item \name{} effectively handles local memory contention and bandwidth interference, achieving up to a \bestimprove{} improvement compared to TPP  (\S\ref{sec:eval-memory-contention}--\S\ref{sec:eval-contention-and-interference}).
    \item \name{}'s real-time adaptation mechanism accurately reallocates resources to prioritize critical applications during runtime workload changes, resulting in 14.9\% and 20.3\% performance improvement compared to TPP and Colloid (\S\ref{sec:eval-real-time}).  

\end{enumerate}

\subsection{Experiment Setup}
\label{sec:eval-setup}
We implement \name{}'s resource management module on Linux kernel v6.6.
\name{}'s user space components are written in C++ with 3000 lines of code. We illustrate \name{}'s effectiveness in providing QoS among 80 workloads (Details on applications are in Appendix~\ref{app:real-apps}). We also compare \name{} with TPP~\cite{tpp} and Colloid~\cite{colloid}, two state-of-the-art open-source tiered memory systems.
We run each experiment five times and report the average measurements.

\subsection{SLO Compliance}
\label{sec:eval-slo-compliance}

\begin{figure}[!t]
    \centering
    \subfloat[SLO compliance for Redis]{
        \label{fig:eval-slo-compliance-lat}
        \includegraphics[width=2.85in]{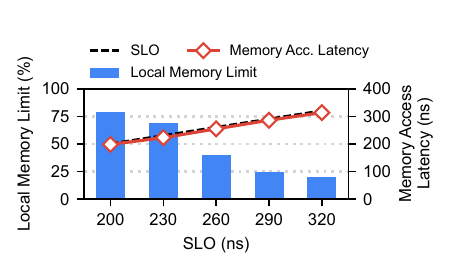}
    } \\
    \subfloat[SLO compliance for llama.cpp]{
        \label{fig:eval-slo-compliance-bw}
        \includegraphics[width=3.2in]{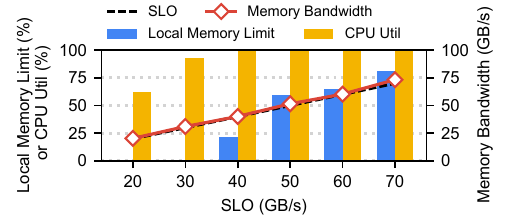}
    }
    \captionsetup{width=\columnwidth}
    \caption{\name{} provides SLO compliance on both memory access latency (for Redis) and bandwidth (for \llama{}).}%
    \label{fig:eval-slo-compliance}
\end{figure}
We start by evaluating how closely \name{} tracks SLOs of both types of applications.
We first look at simple scenarios where a single application is running and leave more complicated multi-tenant settings in the follow-up subsections.
Figure~\ref{fig:eval-slo-compliance-lat} records the achieved memory access latency of Redis at different SLOs, along with the local memory limit (recorded as the percentage w.r.t its 20GB WSS) \name{} sets.
\name{} is able to closely track the SLO by assigning the right amount of local memory.

\name{} is also good at tracking the SLO of BI applications.
Figure~\ref{fig:eval-slo-compliance-bw} shows \llama{}'s bandwidth performance and its resource allocation by \name{} under different SLOs.
When bandwidth SLO is small (e.g., 30GB/s and below in this experiment), \name{} further decreases its CPU utilization, as merely migrating all memory to CXL memory(by setting the local memory limit to zero) still achieves higher bandwidth than needed.
Note that in this case, \name{} does not limit its migration for inter-tier interference, as there is no other LS application running in the system.

\subsection{Handling Local Memory Contention}
\label{sec:eval-memory-contention}

\begin{figure}[tb]
    \centering
    \includegraphics[width=2.6in]{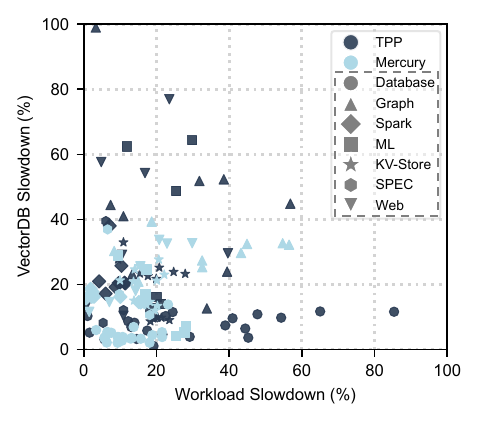}%
    \captionsetup{width=\columnwidth}
    \caption{Comparing \name{} with TPP when handling local memory contention between workloads and VectorDB. }%
    \label{fig:eval-scatter-workload-vectordb}
\end{figure}

Next, we evaluate \name{} in multi-tenant settings to understand its ability to manage local memory contention.
The experiment follows the same workload setup as depicted in Figure \ref{fig:moti-local-dram-contention}, where VectorDB is running as a background application to compete with other workloads.
The results comparing \name{} with TPP are shown in Figure~\ref{fig:eval-scatter-workload-vectordb}.
Notably, most workload coordinates shift towards (0,0), reflecting improved performance.
For Database and KV-store workloads which fail to acquire sufficient local memory when competing with VectorDB under existing solutions, \name{} is able to ensure stable memory allocation, achieving a maximum slowdown reduction of foreground workload from 29\% to 12\%.
In cases where workloads such as Graph, ML, SPEC, Spark, and Web can grab excessive local memory over vectorDB, \name{} manages to maintain VectorDB's SLO and achieves a maximum slowdown reduction of VectorDB from 75\% to 14\%, while minimally affecting the performance of foreground workloads.

\subsection{Handling Memory Bandwidth Interference}
\label{sec:eval-memory-interference}

\begin{figure}[!t]
    \centering
    \includegraphics[width=2.85in]{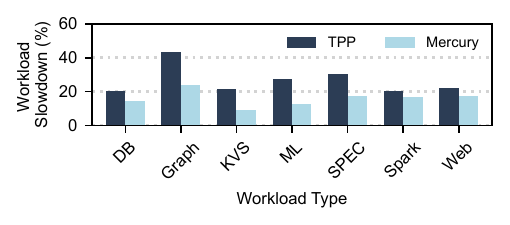}
    \captionsetup{width=\columnwidth}
    \caption{\name{} achieves better performance across all 7 types of workloads for handling memory bandwidth interference.}%
    \label{fig:eval-intra-interference-tpp-mercury}
\end{figure}

In addition to addressing local memory contention, \name{} also tackles memory bandwidth interference as a key focus in its admission control design.
We evaluate \name{}'s effectiveness in managing bandwidth interference across various application combinations.
In these experiments, we configure the working set size (WSS) of applications to eliminate local memory contention.
As illustrated in Figure~\ref{fig:eval-intra-interference-tpp-mercury}, we run \llama{} as a background workload to simulate offline inference scenarios (batch inference), where \llama{}'s priority is low because latency is less critical in this context.
We test 80 foreground workloads spanning 7 diverse types.
For each workload type, \name{} effectively regulates \llama{}'s bandwidth usage, significantly reducing intra-tier interference while avoiding inter-tier interference.
In contrast, TPP fails to meet SLO guarantees, causing foreground workloads to suffer performance degradation due to severe intra-tier interference.
Compared to TPP, \name{} improves the performance of all 80 workloads while having minimal impact on \llama{}'s performance. Notably, for ML workloads, \name{} achieves up to a 40\% performance improvement over TPP.

\subsection{Mixing Two Sources of Unpredictability}
\label{sec:eval-contention-and-interference}

\begin{figure}[tb]
    \centering
    \includegraphics[width=3.3in]{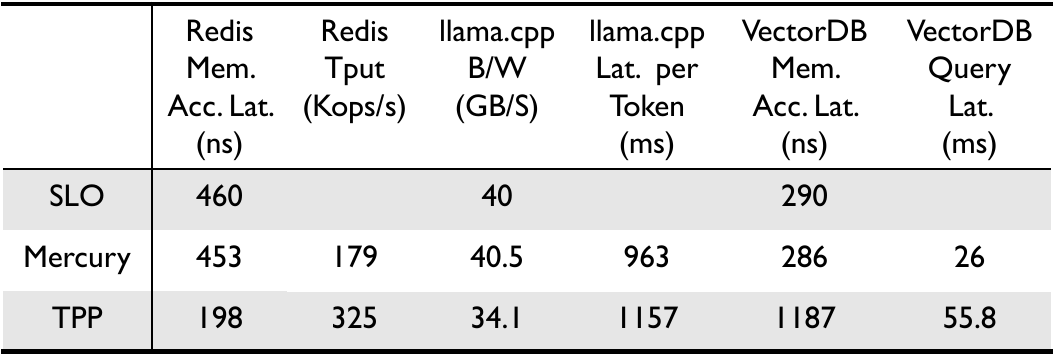}%
    \captionsetup{width=\columnwidth}
    \caption{Comparing \name{} with TPP when handling both local memory contention and memory bandwidth interference among Redis, \llama{}, and VectorDB.}%
    \label{fig:eval-redis-vs-llama-vs-vectordb}
\end{figure}

It is quite likely that local memory contention and memory bandwidth interference appear simultaneously in a real-world setting.
To evaluate how \name{} performs in this case, we deploy Redis, \llama{}, and VectorDB all together with a local memory capacity of 40GB.
The WSS for Redis, \llama{}, and VectorDB is 40GB, 40GB, and 20GB, respectively.
Figure~\ref{fig:eval-redis-vs-llama-vs-vectordb} compares \name{} with TPP on achieved low-level and application-level performance.
TPP allocates almost all local memory to Redis as Redis has the highest memory access frequency among the three applications, and thus most of \llama{}'s and VectorDB's memory are placed on CXL memory.
This allocation satisfies only the SLO of Redis, as \llama{} suffers from insufficient bandwidth, and VectorDB's performance degrades due to interference from \llama{} and not enough local memory. %
In contrast, \name{} satisfies the SLOs for all three applications by allocating the right amount of memory (10GB for Redis, 20GB for VectorDB, and 10GB for \llama{}) and managing \llama{}'s bandwidth to minimize interference, resulting in a 53.4\% performance improvement for VectorDB.

\label{sec:eval-real-time}

\begin{figure}[tb]
    \centering
    {
        \includegraphics[width=3.3in]{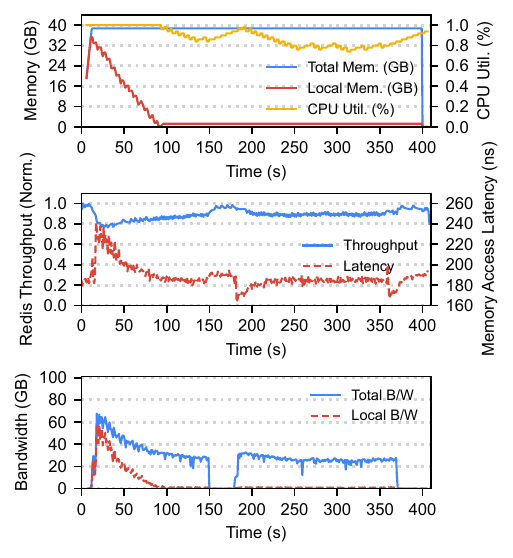}
    } 
    \captionsetup{width=\columnwidth}
    \caption{Performance of Redis and \llama{} under real-time changes using \name. Priority of Redis is higher than \llama{}}%
    \label{fig:real-time-two-app-llama-low}
\end{figure}
\begin{figure}[tb]
    \centering
    {
        \includegraphics[width=3.3in]{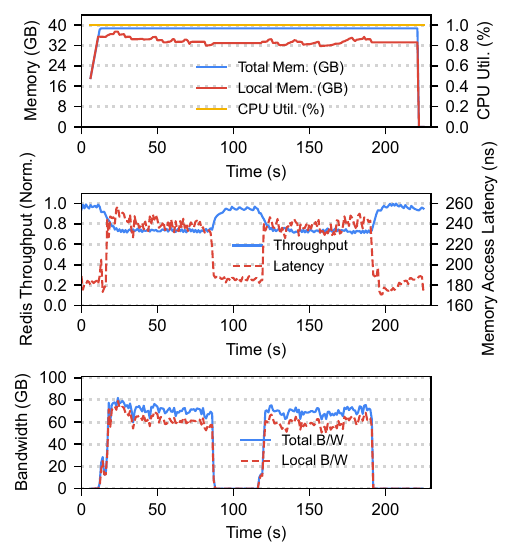}
    } 
    \captionsetup{width=\columnwidth}
    \caption{Performance of Redis and \llama{} under real-time changes using \name{}. Priority of Redis is lower than \llama{}}%
    \label{fig:real-time-two-app-llama-high}
\end{figure}

\begin{figure}[tb]
    \centering
    {
        \label{fig:real-time-low-level}
        \includegraphics[width=3.3in]{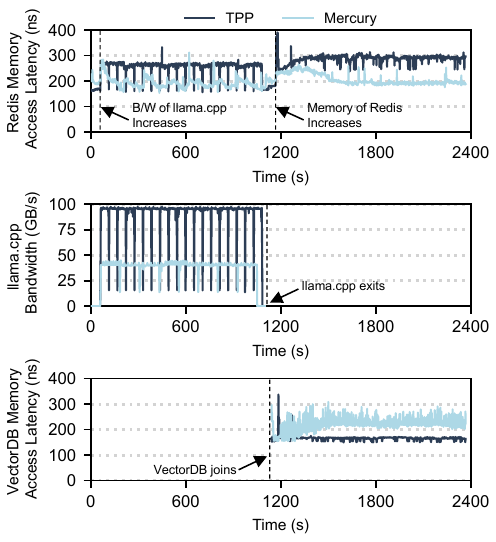}
    } 
    \captionsetup{width=\columnwidth}
    \caption{Performance of Redis, \llama{}, and VectorDB under real-time changes using \name. SLO for Redis/\llama{}/VectorDB is 200ns/180ns/70GBps, with Redis having the highest priority. \llama{}'s bandwidth surges during 60-1100s; Redis's memory usage increases during 1160-2366s.}%
    \label{fig:real-time-low-level}
\end{figure}

\subsection{Real-time Adaptation}
\label{sec:eval-real-time}
\subsubsection{Mitigating Dynamic Bandwidth Interference}
In Section~\ref{sec:compare-colloid}, TPP and Colloid failed to mitigate \llama{}'s bandwidth interference on Redis, leading to high latency spikes and missed SLO. 
An experiment illustrating how \name{} handles QoS in this scenario is shown in Figure \ref{fig:real-time-two-app-llama-low}. %
Initially, \llama{}'s local memory is demoted to remote memory (0--100s) to reduce bandwidth usage. Following this, as all of \llama{}'s memory has been migrated to remote memory, \name{} decreases its CPU utilization and ensure  $thresh_{numa}$ is not exceeded to prevent inter-iteration interference (100--150s). 
Subsequently, \name{} increases \llama{}'s CPU utilization upon detecting that its bandwidth is too low, with no bandwidth interference since no requests are being processed during this period (150--200s). 
Once the second request arrives (after 200s), the system adjusts in a manner similar to the previous behavior.
Compared to TPP and Colloid (in Figure~\ref{fig:moti-colloid-memory-interference}), \name{} improves Redis's average throughput by 14.9\% and 20.3\%, respectively.

Next, we set \llama{}'s priority to be higher and repeat the same experiment to illustrate \name{} can handle various QoS settings.
The results are shown in Figure \ref{fig:real-time-two-app-llama-high}.
In this case, the bandwidth SLO of \llama{} is set at 70 GB/s, which \name{} manages to maintain without being impacted by Redis.
Redis experiences some latency spikes as it now has lower priority, and \name{} prioritize \llama{}'s high bandwidth over Redis performance based on strict priority.

\subsubsection{Long-Running Experiment}
We complete our evaluation with a long running experiment to illustrate how \name{} adapts to different dynamic changes.
Here, we run Redis, \llama{}, and VectorDB together and adjust \llama{}'s bandwidth and Redis's memory usage.
The WSS for Redis, \llama{}, and VectorDB is 30GB, 40GB, and 40GB, respectively, with the local memory capacity constrained at 70 GB.
The priority level of the applications in descending order is Redis, \llama{}, and VectorDB.
We set the SLO for Redis, \llama{}, and VectorDB to be 200ns, 70GB/s, and 180ns.

Initially, Redis and \llama{} launch together without memory contention or bandwidth interference.
When \name{} detects \llama{}'s load arrives at $t=60s$, it quickly mitigates its bandwidth interference to ensure Redis can still maintain its SLO, as Redis has higher priority.
TPP, on the other hand, gives full bandwidth capacity to \llama{}, causing severe interference that violates the SLO of Redis.
After 1100 seconds, \llama{}'s task finishes.
We launch VectorDB and start to gradually increase the load of Redis.
The increase in the memory usage of Redis causes local memory contention.
Under TPP, Redis cannot acquire more memory due to lower access frequency compared to VectorDB. 
In contrast, \name{} reallocates memory from VectorDB to Redis to ensure Redis meets its SLO.
\name{} achieves up to 8.4$\times$ longer SLO satisfaction time for Redis compared to TPP and improved Redis's throughput performance by 33.21\%.

\section{Related Work}
\label{sec:related}
\para{Tiered Memory Systems.}
Many prior solutions\cite{nimble,tmts,hemem,autotiering} consider non-volatile memory (NVM) as the slow memory tier. 
HeMem \cite{hemem} introduces a flexible, per-application memory management policy at the user level. 
AutoTiering \cite{autotiering} improves system utilization by considering access tier and locality without a predefined threshold. 
Diverging from NVM-focused methods, Pond \cite{pond}, TPP \cite{tpp} and Colloid \cite{colloid} explore CXL memory; TPP provides an OS-level, application-transparent mechanism for CXL memory, while Pond develops a predictive model for latency and resource management in a CXL-based memory pool.
Colloid balances every memory access latency to optimize overall system performance.
\name{} decouples memory temperature from application importance, allocating local memory based on SLOs and priority and outperforming previous work that primarily considers applications with hotter pages as more important.

\para{QoS solutions.}
QoS is a full-stack concern addressed across the hardware/software stack. Memshare \cite{memshare} maximizes hit rates and provides isolation in multi-tenant web applications with a log-structured, application-aware design. Aequitas \cite{aequitas} and DiffServ \cite{difserv} prioritize traffic at the network edges, with Aequitas targeting data center environments to ensure latency SLOs for RPCs. TMTS \cite{tmts} uses two metrics to meet SLOs in tiered memory systems, prioritizing LS applications, but it cannot handle memory bandwidth interference. \name{} introduces QoS management that ensures strict priority and fairness across application types.

\para{Interference management.}
Prior systems like Heracles \cite{heracles} and PARTIES \cite{parties} dynamically adjust partitions to handle memory bandwidth interference, with PARTIES providing enhanced isolation for memory capacity and disk bandwidth. MCP \cite{mcp} reduces inter-application interference by mapping data to separate channels, while IMPS \cite{mcp} prioritizes memory non-intensive applications, which can be unfair. The FQ memory scheduler \cite{fairqueuing} prioritizes memory requests by earliest virtual finish-time. ASM \cite{asm} minimizes memory interference by periodically giving each application's requests the highest priority. 
However, these systems do not address inter-tier interference. %
\name{} is the first to combine local memory limits and CPU utilization to avoid both intra-tier and inter-tier bandwidth interference.

\para{Disaggregated Memory.}
Memory disaggregation exposes capacity available in remote hosts as a shared memory pool. Recent RDMA-based disaggregated memory solutions \cite{rdma-1,semeru,rdma-2, infiniswap} face significantly higher latency than CXL memory \cite{direct_cxl}. Memory management in these systems is orthogonal to \name{}; one can use both CXL- and network-enabled memory tiers and apply \name{} to manage tiered memory.

\section{Conclusion}
Tiered memory systems provides higher memory capacity to allow more memory-intensive applications to be deployed.
However, existing tiered memory systems focus on optimizing a single application, and cannot provide QoS guarantees when multiple applications sharing memory resource.
We present the design and implementation of \name{}, a QoS-aware tiered memory system to provide predictable performance for coexisting memory-intensive workloads.
\name{} provides application-level resource management by enabling per-tier page reclamation inside Linux kernel.
By designing a novel admission control algorithm, \name{} maximizes local memory utilization while mitigate both intra-tier and inter-tier memory bandwidth interference.
\name{} outperforms the state-of-the-art solution when handling local memory contention, memory interference, and dynamic workload changes with significant performance improvement.

\clearpage

\bibliographystyle{plain}                  %
\def\UrlBreaks{\do\/\do-}
\bibliography{reference}

\begin{thebibliography}{10}

\bibitem{hibench}
{HiBench Suite.}
\newblock \url{https://github.com/Intelbigdata/HiBench}.

\bibitem{postgre}
{PostgreSQL Database Management System}.
\newblock \url{https://www.postgresql.org/}.

\bibitem{spec}
{SPEC CPU 2017}.
\newblock \url{https://www.spec.org/cpu2017/}.

\bibitem{tpc-c}
{TPC Benchmark C (TPC-C)}.
\newblock \url{http://www.tpc.org/tpcc/}.

\bibitem{tpc-h}
{TPC Benchmark H (TPC-H)}.
\newblock \url{https://www.tpc.org/tpch/}.

\bibitem{linux-interleaving}
{Weighted interleaving for memory tiering}.
\newblock \url{https://lwn.net/Articles/948037/}.

\bibitem{numa-balancing}
Numa balancing.
\newblock \url{https://mirrors.edge.kernel.org/pub/linux/kernel/people/andrea/autonuma/autonuma_bench-20120530.pdf}, 2012.

\bibitem{cgroup}
Linux cgroups.
\newblock \url{https://docs.kernel.org/admin-guide/cgroup-v2.html}, 2015.

\bibitem{pqos}
Intel platform qos technologies.
\newblock \url{https://wiki.xenproject.org/wiki/Intel_Platform_QoS_Technologies}, 2018.

\bibitem{pebs}
Pebs. intel 64 and ia-32 architectures software developer's manual.
\newblock \url{https://software.intel.com/articles/intel-sdm/}, 2023.

\bibitem{cxl}
Compute express link (cxl).
\newblock \url{https://www.computeexpresslink.org/}, 2024.

\bibitem{llama}
Inference of meta's llama model (and others) in pure c/c++.
\newblock \url{https://github.com/ggerganov/llama.cpp/}, 2024.

\bibitem{memcached}
memcached.
\newblock \url{https://memcached.org/}, 2024.

\bibitem{redis}
Redis.
\newblock \url{https://github.com/redis/redis/}, 2024.

\bibitem{thermostat}
Neha Agarwal and Thomas~F. Wenisch.
\newblock Thermostat: Application-transparent page management for two-tiered main memory.
\newblock In {\em ASPLOS}, 2017.

\bibitem{uprof}
AMD.
\newblock {$\mu$Prof User Guide}.
\newblock 2024.

\bibitem{gap-benchmark}
Scott Beamer, Krste Asanovi{\'c}, and David Patterson.
\newblock The gap benchmark suite.
\newblock {\em arXiv preprint arXiv:1508.03619}, 2015.

\bibitem{difserv}
S.~Blake, D.~Black, M.~Carlson, E.~Davies, Z.~Wang, , and W.~Weiss.
\newblock Rfc2475: An architecture for differentiated service.
\newblock In {\em IETF}, 1998.

\bibitem{rdma-1}
Irina Calciu, M~Talha Imran, Ivan Puddu, Sanidhya Kashyap, Hasan~Al Maruf, Onur Mutlu, and Aasheesh Kolli.
\newblock Rethinking software runtimes for disaggregated memory.
\newblock In {\em Proceedings of the 26th ACM International Conference on Architectural Support for Programming Languages and Operating Systems}, pages 79--92, 2021.

\bibitem{faster}
Badrish Chandramouli, Guna Prasaad, Donald Kossmann, Justin Levandoski, James Hunter, and Mike Barnett.
\newblock Faster: A concurrent key-value store with in-place updates.
\newblock In {\em Proceedings of the 2018 International Conference on Management of Data}, pages 275--290, 2018.

\bibitem{atlas}
Lei Chen, Shi Liu, Chenxi Wang, Haoran Ma, Yifan Qiao, Zhe Wang, Chenggang Wu, Youyou Lu, Xiaobing Feng, Huimin Cui, et~al.
\newblock A tale of two paths: Toward a hybrid data plane for efficient far-memory applications.
\newblock In {\em 18th USENIX Symposium on Operating Systems Design and Implementation (OSDI 24)}, pages 77--95, 2024.

\bibitem{parties}
Shuang Chen, Christina Delimitrou, and José~F. Martínez.
\newblock Parties: Qos-aware resource partitioning for multiple interactive services.
\newblock In {\em ASPLOS}, 2019.

\bibitem{memshare}
Asaf Cidon, Daniel Rushton, Stephen~M. Rumble, and Ryan Stutsman.
\newblock Memshare: a dynamic multi-tenant key-value cache.
\newblock In {\em ATC}, 2017.

\bibitem{ycsb}
Brian~F Cooper, Adam Silberstein, Erwin Tam, Raghu Ramakrishnan, and Russell Sears.
\newblock Benchmarking cloud serving systems with ycsb.
\newblock In {\em Proceedings of the 1st ACM symposium on Cloud computing}, pages 143--154, 2010.

\bibitem{faiss}
Matthijs Douze, Alexandr Guzhva, Chengqi Deng, Jeff Johnson, Gergely Szilvasy, Pierre-Emmanuel Mazaré, Maria Lomeli, Lucas Hosseini, and Hervé Jégou.
\newblock The faiss library.
\newblock 2024.

\bibitem{ibs}
Paul~J. Drongowski.
\newblock Instruction-based sampling: A new performance analysis technique for amd family 10h processors.
\newblock 2007.

\bibitem{tmts}
Padmapriya Duraisamy, Wei Xu, Scott Hare, Ravi Rajwar, David Culler, Zhiyi Xu, Jianing Fan, Chris Kennelly, Bill McCloskey, Danijela Mijailovic, Brian Morris, Chiranjit Mukherjee, Jingliang Ren, Greg Thelen, Paul Turner, Carlos Villavieja, Parthasarathy Ranganathan, and Amin Vahdat.
\newblock Towards an adaptable systems architecture for memory tiering at warehouse-scale.
\newblock In {\em ASPLOS}, 2023.

\bibitem{rdma-2}
Yixiao Gao, Qiang Li, Lingbo Tang, Yongqing Xi, Pengcheng Zhang, Wenwen Peng, Bo~Li, Yaohui Wu, Shaozong Liu, Lei Yan, et~al.
\newblock When cloud storage meets {RDMA}.
\newblock In {\em 18th USENIX Symposium on Networked Systems Design and Implementation (NSDI 21)}, pages 519--533, 2021.

\bibitem{direct_cxl}
Donghyun Gouk, Sangwon Lee, Miryeong Kwon, and Myoungsoo Jung.
\newblock Direct access, {High-Performance} memory disaggregation with {DirectCXL}.
\newblock In {\em 2022 USENIX Annual Technical Conference (USENIX ATC 22)}, pages 287--294, 2022.

\bibitem{infiniswap}
Juncheng Gu, Youngmoon Lee, Yiwen Zhang, Mosharaf Chowdhury, and Kang~G. Shin.
\newblock Efficient memory disaggregation with {Infiniswap}.
\newblock In {\em NSDI}, 2017.

\bibitem{dlrm-benchmark-1}
Rishabh Jain, Scott Cheng, Vishwas Kalagi, Vrushabh Sanghavi, Samvit Kaul, Meena Arunachalam, Kiwan Maeng, Adwait Jog, Anand Sivasubramaniam, Mahmut~Taylan Kandemir, et~al.
\newblock Optimizing cpu performance for recommendation systems at-scale.
\newblock In {\em Proceedings of the 50th Annual International Symposium on Computer Architecture}, pages 1--15, 2023.

\bibitem{HeteroOS}
Sudarsun Kannan, Ada Gavrilovska, Vishal Gupta, and Karsten Schwan.
\newblock Heteroos: Os design for heterogeneous memory management in datacenter.
\newblock In {\em ISCA}, 2017.

\bibitem{autotiering}
Jonghyeon Kim, Wonkyo Choe, and Jeongseob Ahn.
\newblock Exploring the design space of page management for multi-tiered memory systems.
\newblock In {\em ATC}, 2021.

\bibitem{pond}
Huaicheng Li, Daniel~S. Berger, Stanko Novakovic, Lisa Hsu, Dan Ernst, Pantea Zardoshti, Monish Shah, Samir Rajadnya, Scott Lee, Ishwar Agarwal, Mark~D. Hill, Marcus Fontoura, and Ricardo Bianchini.
\newblock Pond: Cxl-based memory pooling systems for cloud platforms.
\newblock In {\em ASPLOS}, 2023.

\bibitem{heracles}
David Lo, Liqun Cheng, Rama Govindaraju, Parthasarathy Ranganathan, and Christos Kozyrakis.
\newblock Heracles: Improving resource efficiency at scale.
\newblock In {\em ISCA}, 2015.

\bibitem{tpp}
Hasan~Al Maruf, Hao Wang, Abhishek Dhanotia, Johannes Weiner, Niket Agarwal, Pallab Bhattacharya, Chris Petersen, Mosharaf Chowdhury, Shobhit Kanaujia, and Prakash Chauhan.
\newblock {TPP}: Transparent page placement for {CXL}-enabled tiered memory.
\newblock In {\em ASPLOS}, 2023.

\bibitem{mcp}
Sai~Prashanth Muralidhara, Lavanya Subramanian, Onur Mutlu, Mahmut Kandemir, and Thomas Moscibroda.
\newblock Reducing memory interference in multicore systems via application-aware memory channel partitioning.
\newblock In {\em 2011 44th Annual IEEE/ACM International Symposium on Microarchitecture (MICRO)}, pages 374--385, 2011.

\bibitem{dlrm}
Maxim Naumov, Dheevatsa Mudigere, Hao{-}Jun~Michael Shi, Jianyu Huang, Narayanan Sundaraman, Jongsoo Park, Xiaodong Wang, Udit Gupta, Carole{-}Jean Wu, Alisson~G. Azzolini, Dmytro Dzhulgakov, Andrey Mallevich, Ilia Cherniavskii, Yinghai Lu, Raghuraman Krishnamoorthi, Ansha Yu, Volodymyr Kondratenko, Stephanie Pereira, Xianjie Chen, Wenlin Chen, Vijay Rao, Bill Jia, Liang Xiong, and Misha Smelyanskiy.
\newblock Deep learning recommendation model for personalization and recommendation systems.
\newblock {\em CoRR}, abs/1906.00091, 2019.

\bibitem{fairqueuing}
Kyle~J. Nesbit, Nidhi Aggarwal, James Laudon, and James~E. Smith.
\newblock Fair queuing memory systems.
\newblock In {\em 2006 39th Annual IEEE/ACM International Symposium on Microarchitecture (MICRO'06)}, pages 208--222, 2006.

\bibitem{mba-intel}
Khang~T Nguyen.
\newblock Introduction to memory bandwidth monitoring in the {Intel Xeon} processor e5 v4 family.
\newblock 2016.

\bibitem{tmc}
Yuanjiang Ni, Pankaj Mehra, Ethan Miller, and Heiner Litz.
\newblock Tmc: Near-optimal resource allocation for tiered-memory systems.
\newblock In {\em SoCC}, 2023.

\bibitem{renaissance}
Aleksandar Prokopec, Andrea Ros{\`a}, David Leopoldseder, Gilles Duboscq, Petr T\r{u}ma, Martin Studener, Lubom{\'\i}r Bulej, Yudi Zheng, Alex Villaz{\'o}n, Doug Simon, et~al.
\newblock Renaissance: A modern benchmark suite for parallel applications on the jvm.
\newblock In {\em SPLASH}, 2019.

\bibitem{hemem}
Amanda Raybuck, Tim Stamler, Wei Zhang, Mattan Erez, and Simon Peter.
\newblock Hemem: Scalable tiered memory management for big data applications and real nvm.
\newblock In {\em ASPLOS}, pages 392--407, 2021.

\bibitem{vtmm}
Sai Sha, Chuandong Li, Yingwei Luo, Xiaolin Wang, and Zhenlin Wang.
\newblock vtmm: Tiered memory management for virtual machines.
\newblock In {\em EuroSys}, 2023.

\bibitem{asm}
Lavanya Subramanian, Vivek Seshadri, Arnab Ghosh, Samira Khan, and Onur Mutlu.
\newblock The application slowdown model: Quantifying and controlling the impact of inter-application interference at shared caches and main memory.
\newblock In {\em MICRO}, 2015.

\bibitem{silo}
Stephen Tu, Wenting Zheng, Eddie Kohler, Barbara Liskov, and Samuel Madden.
\newblock Speedy transactions in multicore in-memory databases.
\newblock In {\em Proceedings of the Twenty-Fourth ACM Symposium on Operating Systems Principles}, pages 18--32, 2013.

\bibitem{colloid}
Midhul Vuppalapati and Rachit Agarwal.
\newblock Tiered memory management: Access latency is the key!
\newblock In {\em SOSP}, 2024.

\bibitem{semeru}
Chenxi Wang, Haoran Ma, Shi Liu, Yuanqi Li, Zhenyuan Ruan, Khanh Nguyen, Michael~D Bond, Ravi Netravali, Miryung Kim, and Guoqing~Harry Xu.
\newblock Semeru: A {Memory-Disaggregated} managed runtime.
\newblock In {\em 14th USENIX Symposium on Operating Systems Design and Implementation (OSDI 20)}, pages 261--280, 2020.

\bibitem{canvas}
Chenxi Wang, Yifan Qiao, Haoran Ma, Shi Liu, Wenguang Chen, Ravi Netravali, Miryung Kim, and Guoqing~Harry Xu.
\newblock Canvas: Isolated and adaptive swapping for {Multi-Applications} on remote memory.
\newblock In {\em 20th USENIX Symposium on Networked Systems Design and Implementation (NSDI 23)}, pages 161--179, 2023.

\bibitem{nimble}
Zi~Yan, Daniel Lustig, David Nellans, and Abhishek Bhattacharjee.
\newblock Nimble page management for tiered memory systems.
\newblock In {\em ASPLOS}, 2023.

\bibitem{mt2}
Jifei Yi, Benchao Dong, Mingkai Dong, Ruizhe Tong, and Haibo Chen.
\newblock {MT$^2$: Memory Bandwidth Regulation on Hybrid NVM/DRAM Platforms}.
\newblock In {\em 20th USENIX Conference on File and Storage Technologies (FAST 22)}, pages 199--216, 2022.

\bibitem{spark}
Matei Zaharia, Mosharaf Chowdhury, Tathagata Das, Ankur Dave, Justin Ma, Murphy McCauly, Michael~J Franklin, Scott Shenker, and Ion Stoica.
\newblock Resilient distributed datasets: A {Fault-Tolerant} abstraction for {In-Memory} cluster computing.
\newblock In {\em 9th USENIX symposium on networked systems design and implementation (NSDI 12)}, pages 15--28, 2012.

\bibitem{aequitas}
Yiwen Zhang, Gautam Kumar, Nandita Dukkipati, Xian Wu, Priyaranjan Jha, Mosharaf Chowdhury, and Amin Vahdat.
\newblock {Aequitas}: Admission control for performance-critical rpcs in datacenters.
\newblock In {\em SIGCOMM}, 2022.

\bibitem{memstrata}
Yuhong Zhong, Daniel~S Berger, Carl Waldspurger, Ishwar Agarwal, Rajat Agarwal, Frank Hady, Karthik Kumar, Mark~D Hill, Mosharaf Chowdhury, and Asaf Cidon.
\newblock Managing memory tiers with cxl in virtualized environments.
\newblock In {\em Symposium on Operating Systems Design and Implementation}, 2024.

\end{thebibliography}
\clearpage
\appendix
\nobalance
\section{Workloads}
\label{app:real-apps}

We use a total of 80 real-world applications: 
\begin{denseitemize}
    \item \textbf{Database:} TPC-C~\cite{tpc-c} on Silo~\cite{silo} and TPC-H~\cite{tpc-h} on PostgreSQL~\cite{postgre} and Faiss~\cite{faiss}
    \item \textbf{Machine learning (ML):} DLRM benchmark~\cite{dlrm-benchmark-1,dlrm} and llama.cpp~\cite{llama}
    \item \textbf{Key-value (KV) store:} YCSB~\cite{ycsb} on FASTER~\cite{faster}, Redis~\cite{redis}
    \item \textbf{Big data:} HiBench~\cite{hibench} on Spark~\cite{spark}
    \item \textbf{Graph processing:} GAP benchmark~\cite{gap-benchmark}
    \item \textbf{Scientific computing:} SPEC CPU 2017~\cite{spec}
    \item \textbf{Web:} Renaissance~\cite{renaissance}
\end{denseitemize}

\end{document}